\documentclass[prl,aps,twocolumn,preprintnumbers,amssymb,nobibnotes,nofootinbib,raggedbottom,10pt]{revtex4-2}
\usepackage{amsmath}
\usepackage{amssymb}
\usepackage{hyperref,graphicx,color,amsmath,dsfont,titlesec,shuffle,tikz}
\usepackage{upgreek}
\usepackage{mathtools}
\usepackage{caption}
\usepackage{subcaption}
\usepackage{hyperref}
\usepackage[most]{tcolorbox}
\titleformat{\section}{\centering\normalsize\normalfont\bf}{\thesection}{0em}{}
\hypersetup{pdftitle={},pdfcreator={},linkcolor=[rgb]{0.15,0.35,0.75},colorlinks=true,citecolor=[rgb]{0.675,0,0.2},urlcolor=[rgb]{0.15,0.35,0.65}}
\definecolor{nmhv}{rgb}{0.95,0.55,0.55}
\definecolor{mhvblue}{rgb}{0.6,0.6,0.7765}
\definecolor{ampgrey}{rgb}{0.9,0.9,0.9}
\definecolor{hblue}{rgb}{0,0,0.575}
\definecolor{hred}{rgb}{0.575,0.0,0.225}
\definecolor{hgreen}{rgb}{0.0,0.4,0.2}
\definecolor{hteal}{rgb}{0.0,0.545,0.7451}
\definecolor{perm}{rgb}{0.1,0.45,0.85}
\definecolor{unord}{rgb}{0,0,0}
\definecolor{ord}{rgb}{0,0,0.575}
\definecolor{anchorLeg}{rgb}{0.575,0.0,0.225}
\definecolor{fRed}{rgb}{0.48,0.02824,0.18824}

\makeatletter 
    
\renewcommand\onecolumngrid{
\do@columngrid{one}{\@ne}
\def\set@footnotewidth{\onecolumngrid}
\def\footnoterule{\kern-6pt\hrule width 1.5in\kern6pt}
}

\renewcommand\twocolumngrid{
        \def\footnoterule{
        \dimen@\skip\footins\divide\dimen@\thr@@
        \kern-\dimen@\hrule width.5in\kern\dimen@}
        \do@columngrid{mlt}{\tw@}
}

\makeatother

\newcommand{\tr}{\text{Tr}}
\newcommand{\trp}{\smash{\tr(\phi^3)}}
\newcommand{\dd}{\mathrm{d}}

\newcommand{\Tan}[1]{\tan(\pi\alpha' #1)}
\newcommand{\Sin}[1]{\sin(\pi\alpha' #1)}

\begin{document}
\title {Positive Geometry for Stringy Scalar Amplitudes}
\author{Christoph Bartsch} 
\author{Karol Kampf}
\author{David Podiv\'in}
\author{Jonah Stalknecht}

\affiliation{Institute for Particle and Nuclear Physics, Charles University, Prague, Czech Republic}

\begin{abstract}
    \noindent We introduce a new positive geometry, the associahedral grid, which provides a geometric realization of the inverse string theory KLT kernel. It captures the full $\alpha'$-dependence of stringified amplitudes for bi-adjoint scalar $\phi^3$ theory, pions in the NLSM, and their mixed $\phi$/$\pi$ amplitudes, reducing to the corresponding field theory amplitudes in the $\alpha'\to 0$ limit. Our results demonstrate how positive geometries can be utilized beyond rational functions to capture stringy features of amplitudes, such as an infinite resonance structure. The kinematic $\delta$-shift, recently proposed to relate field theory $\mathrm{Tr}(\phi^3)$ and NLSM pion amplitudes, naturally emerges as the leading contribution to the stringy geometry. We show how the connection between $\mathrm{Tr}(\phi^3)$ and NLSM can be geometrized using the associahedral grid.
\end{abstract}

\maketitle

\vspace{-0.3cm}

\subparagraph{I. Introduction.}

Since the discovery of the amplituhedron for $\mathcal{N}{=}4$ supersymmetric Yang-Mills theory \cite{Arkani-Hamed:2013jha}, positive geometry \cite{Arkani-Hamed:2017tmz,Brown:2025jjg} has repeatedly proven effective in revealing unexpected structures in field theory scattering amplitudes \cite{Arkani-Hamed:2013kca,Arkani-Hamed:2014dca,Banerjee:2018tun,Arkani-Hamed:2018rsk,Damgaard:2019ztj,Herrmann:2020qlt,Damgaard:2020eox,Lukowski:2021fkf,Arkani-Hamed:2021iya,He:2021llb,Huang:2021jlh,Arkani-Hamed:2023epq,Brown:2023mqi,Trnka:2020dxl,Paranjape:2023qsq,Koefler:2024pzv,Lagares:2024epo,Henn:2023pkc,He:2023exb,Ferro:2015grk,Franco:2014csa,Lukowski:2022fwz,Ferro:2020lgp,Damgaard:2021qbi,Ferro:2018vpf,Arkani-Hamed:2017vfh,Karp:2017ouj,Even-Zohar:2021sec,Even-Zohar:2025ydi,Even-Zohar:2023del}, correlators \cite{Eden:2017fow,He:2024xed,He:2025rza}, and cosmology \cite{Arkani-Hamed:2017fdk, Arkani-Hamed:2024jbp, Capuano:2025ehm, Glew:2025otn, De:2025bmf, Figueiredo:2025daa,Benincasa:2024leu,Glew:2025ugf,Glew:2025ypb}. Geometric and combinatorial concepts also underlie the simplest theory of colored scalars, the $\trp$ theory, leading to recent advances such as the 'surfaceology' framework \cite{Arkani-Hamed:2023lbd,Arkani-Hamed:2023mvg,Arkani-Hamed:2024pzc,De:2024wsy}, `hidden zeros' \cite{Arkani-Hamed:2023swr,Bartsch:2024amu,Jones:2025rbv,Cao:2024gln,Li:2024qfp,Li:2024bwq,Rodina:2024yfc,Backus:2025hpn}, and $\delta$-shift relations to pions in the Non-linear sigma model (NLSM) and gluon amplitudes \cite{Arkani-Hamed:2024yvu,Arkani-Hamed:2024nhp,Arkani-Hamed:2024vna,Paranjape:2025wjk}.

The $\trp$ theory is contained in the bi-adjoint scalar (BAS) model, whose amplitudes are encoded by canonical differential forms defined on a positive geometry directly in kinematic space known as the Arkani-Hamed--Bai--He--Yan (ABHY) associahedron \cite{Arkani-Hamed:2017mur}. BAS theory also features prominently in string-inspired approaches, including the Cachazo--He--Yuan formalism \cite{Cachazo:2013gna,Cachazo:2013hca,Cachazo:2013iea} and the Kawai--Lewellen--Tye (KLT) double copy \cite{Bern:2019prr}. The latter suggests a canonical `stringification' of BAS amplitudes via the inverse string theory KLT kernel \cite{Mizera:2016jhj},
\begin{align}\label{defKLTinv}
	m^{\alpha'}_n \equiv (\mathcal{S}_{\alpha'})^{-1}.
\end{align}
The KLT kernel $\mathcal{S}_{\alpha'}$ is a central object in string theory, arising from the double-copy structure of closed and open string amplitudes \cite{Kawai:1985xq,Bjerrum-Bohr:2010pnr}. 

In this letter we extend the ABHY associahedron for field theory BAS amplitudes into a stringy positive geometry for the inverse KLT kernel \eqref{defKLTinv}. This is accomplished by introducing the \textit{associahedral grid}, which can be pictured as an infinite array of associahedra capturing the periodic, string-like resonance structure of $m_n^{\alpha'}$. The associahedral grid provides a first example for a stringy positive geometry capturing the analytic $\alpha'$-dependence of physical quantities, going beyond those described by rational functions.

A particular strength of the infinite grid picture is that it geometrizes the connection between $\trp$ amplitudes and pions in the NLSM recently observed in the context of the $\delta$-shift and the $\alpha'$-shift \cite{Bartsch:2025loa} for $m_n^{\alpha'}$ in particular. In \cite{Bartsch:2025loa} it was shown that the inverse KLT kernel unifies amplitudes of cubic scalars, pions, and mixed $\phi/\pi$ amplitudes into a common stringy framework, relating them via kinematic $\alpha'$-shifts. This allows us to also derive geometries for stringy pions and mixed amplitudes. The $\alpha'$-shifts act on the associahedral grid of the inverse KLT kernel $m_n^{\alpha'}$ as rescalings and translations. Geometrically, this selects some associahedral \emph{subgrid} whose canonical form then encodes pions and mixed amplitudes. This extends the applicability of positive geometries to a new class of theories with non-trivial kinematic numerators and UV poles.

\subparagraph{IA. Review: The ABHY Associahedron.}

The ABHY associahedron $\mathcal{A}_n$ is a polytope embedded in the $n(n-3)/2$-dimensional kinematic space of massless $n$-particle scattering. The construction depends on a choice of kinematic basis containing some set of $(n-3)$ planar Mandelstam variables
\begin{align}\label{eq:defX}
	X_{ij} = (p_i+p_{i+1}+\dots + p_{j-1})^2, \hspace{0.5cm} 1\le i < j \le n.
\end{align}
We take the remaining $(n-2)(n-3)/2$ basis elements from the set of non-planar variables
\begin{align}
    c_{ij}\coloneqq X_{ij}+X_{i+1j+1}-X_{ij+1}-X_{i+1j}=-2p_i\cdot p_j,
\end{align}
where the indices $i,j$ are (cyclically) non-adjacent. Throughout this letter we employ a `standard' basis containing $\{X_{1i}\}_{i=3}^{n-1}$, together with the set of all $c_{ij}$ variables with $i,j\neq n$ such that any $X_{ab}$ can be uniquely expressed in terms of this basis. We then define the ABHY associahedron $\mathcal{A}_n$ to be the polytope cut out by all inequalities $X_{ab}\geq 0$, together with the hypersurface constraints that all $c_{ij}$s in our kinematic basis are positive constants.

The ABHY associahedron fits within the framework of \emph{positive geometries} \cite{Arkani-Hamed:2017tmz}, where it is attributed a specific \emph{canonical form}. The canonical form $\Omega(\mathcal{A}_n)$ is defined to be a top-form with logarithmic singularities at the boundaries of the geometry. When a residue at one of these singularities is taken, the resulting differential form is then required to be the canonical form of the boundary. A zero-dimensional positive geometry is a point, and its canonical form is required to be $\pm 1$. The most important example for our discussion is the line segment 
\begin{align}\label{eq:Omega-line-segment}
    \Omega\big(\{a\leq x \leq b\}\big) = \dd \log\frac{x-a}{x-b},
\end{align}
which will serve as a building block for all geometries we consider. Notably, the 4-point ABHY associahedron is the line segment
\begin{align}
    \mathcal{A}_4 = \lbrace 0 \le X_{13} \le c_{13} \rbrace.
\end{align}
Its canonical form
\begin{align}\label{trp4pt}
\Omega(\mathcal{A}_4) =\! \left( \frac{1}{X_{13}} + \frac{1}{X_{24}} \right)\dd X_{13}
\end{align}
encodes the 4-point $\trp$ amplitude. This holds for $n>4$ where the canonical form of the ABHY associahedron
\begin{align}
    \Omega(\mathcal{A}_n)= m_n \dd^{n-3}X,
\end{align}
captures the $n$-point amplitude $m_n$ of $\trp$ theory.
\subparagraph{IB. Review: Inverse Stringy KLT Kernel.}
We briefly review some pertinent features of the matrix elements $m_n^{\alpha'}(\sigma | \rho)$. In \cite{Mizera:2016jhj,Mizera:2017cqs} it was shown that the matrix elements are simple trigonometric functions of the external kinematics.

Using $n$-point planar Mandelstam invariants \eqref{eq:defX}, some explicit examples for \eqref{defKLTinv} include the three-point function $m_3^{\alpha'} (\mathds{1}|\mathds{1}) = 1$ as well as
\begin{align}
     m_4^{\alpha'}\!\!(\mathds{1}|\mathds{1}) &= \frac{1}{\Tan{X_{13}}}+\frac{1}{\Tan{X_{24}}}, \label{invKLTex}\\
     m_5^{\alpha'}\!\!(\mathds{1}|\mathds{1}) &= \left(\!\frac{1}{\Tan{X_{13}}\Tan{X_{14}}} + \text{cyc.}\!\right) \!+\!1, \nonumber
\end{align}
at four and five points. 

A notable difference compared to BAS field theory amplitudes is the presence of an infinite tower of odd-point contact interactions \cite{Mizera:2016jhj}, as illustrated by the five-point function in \eqref{invKLTex}. Another characteristic feature of the stringy matrix elements $m_n^{\alpha'}\!(\sigma | \rho)$ is their periodic pole structure. They have simple poles when any
\begin{align}\label{periodPoles}
    X_{ij} = k/\alpha', \hspace{0.5cm} k\in\mathbb{Z},
\end{align}
on all of which they consistently factorize into a product of two lower-point matrix elements.

In the above we have given examples of \textit{diagonal} matrix elements $m_n^{\alpha'}\!(\mathds{1}|\mathds{1})$ where $\mathds{1}=\lbrace 1\dots n\rbrace$ denotes the identity permutation of $n$ particle labels. We will also consider geometries for \textit{off-diagonal} matrix elements $m_n^{\alpha'}\!(\sigma|\rho)$ with $\sigma \neq \rho$, for example
\begin{align}
    m_3^{\alpha'}\!\!(\mathds{1}|132) &=  -1, \hspace{0.3cm} m_4^{\alpha'}\!\!(\mathds{1}|1243) = \frac{-1}{\Sin{X_{13}}}, \label{exKLToffdiag} \\
    \hspace{-0.15cm} m_5^{\alpha'}\!\!(\mathds{1}|13452) &{=}\frac{1}{\Sin{X_{13}}}\!\!\left(\!\!\frac{1}{\Tan{X_{14}}}{+}\frac{1}{\Tan{X_{35}}}\!\!\right)\!\! \nonumber.
\end{align}
More generally, off-diagonal matrix elements are always given by a product of factors $\sin(\pi\alpha' X_{ij})^{-1}$ and lower-point diagonal matrix elements \cite{Mizera:2016jhj}.

In the field theory limit $\alpha' \to 0$ the stringy matrix elements \eqref{defKLTinv} reduce to amplitudes $m_n(\sigma | \rho)$ of the BAS theory,
\begin{align} \label{alpha0limBAS}
    m_n^{\alpha'}\!(\sigma | \rho ) = (\pi\alpha')^{3-n}\big(m_n (\sigma|\rho) + \mathcal{O}(\alpha')\big).
\end{align}
We will denote the diagonal elements simply as $m_n^{\alpha'}\equiv m_n^{\alpha'}(\mathds{1}|\mathds{1})$, and $m_n\equiv m_n(\mathds{1}|\mathds{1})$. The latter are identical to the $\trp$ amplitudes $m_n$ discussed earlier.

Given the above properties, any tentative geometry for the stringy matrix elements $m_n^{\alpha'}(\sigma | \rho)$ has to (i) account for the presence of an infinite number of contact terms and (ii) encode the periodic pole structure \eqref{periodPoles}. Additionally, it has to (iii) reduce to the known ABHY associahedron in the limit $\alpha' \to 0$ due to \eqref{alpha0limBAS}.

\begin{figure}
    \centering
    \includegraphics[width=\linewidth]{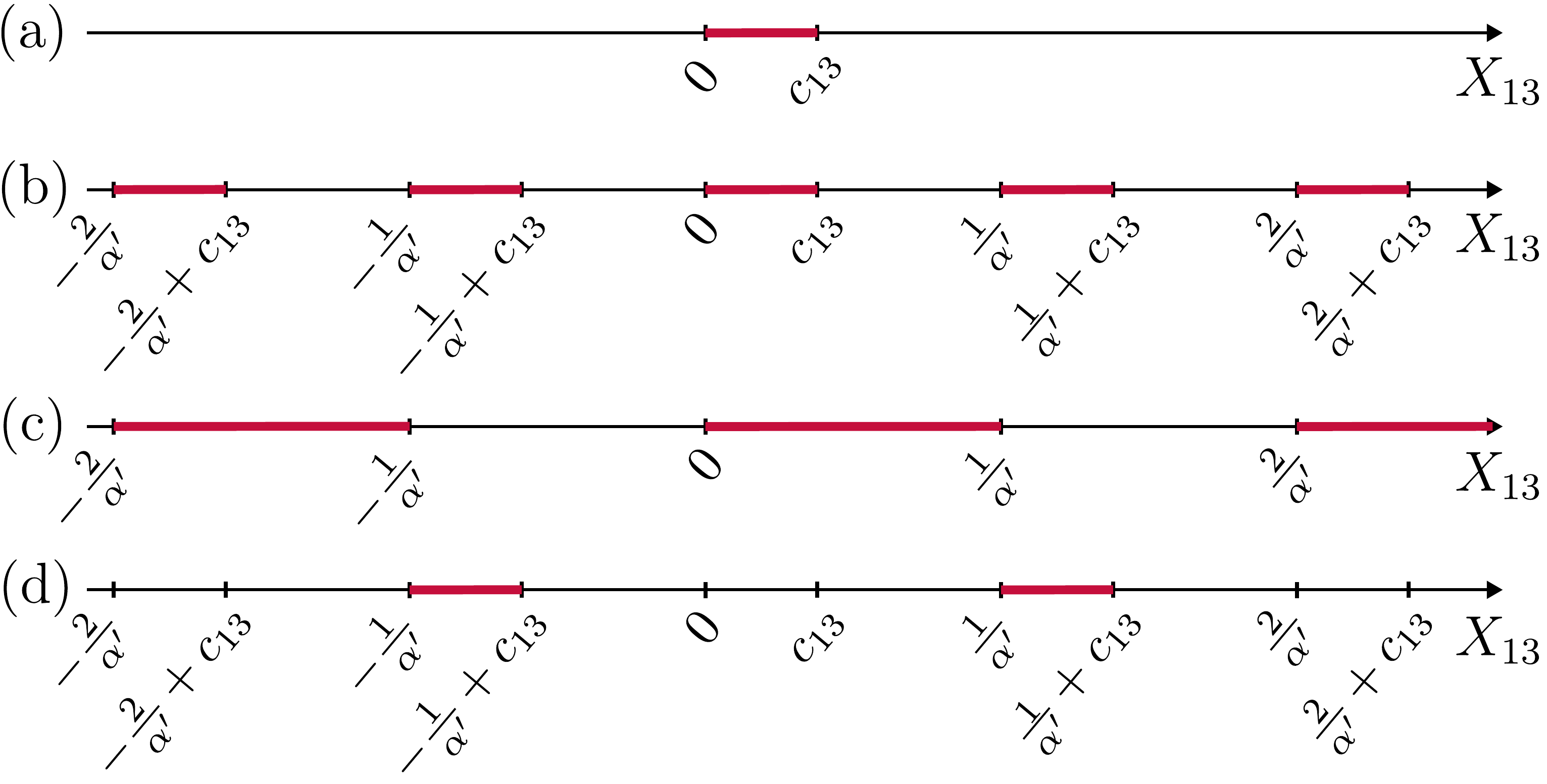}
    \caption{Four-point geometries: (a) $\trp$ amplitude ($\mathcal{A}_4$), (b) diagonal matrix element $m_4^{\alpha'}\!(\mathds{1}|\mathds{1})$ \eqref{invKLTex} ($\mathcal{A}_4^{\alpha'}$), (c) off-diagonal matrix element $m_4^{\alpha'}\!(\mathds{1}|1243)$ \eqref{exKLToffdiag}, (d) stringy NLSM \eqref{NLSM4dform} ($\mathcal{A}_4^{\text{NLSM},\alpha'}$).}
    \label{fig:A4-all}
\end{figure}

\subparagraph{II. Positive Geometry for Diagonal Matrix Elements.}
To construct a geometry for $m_n^{\alpha'}$, we start by showing that the stringy 4-point function \eqref{invKLTex} can be associated to a positive geometry given by an infinite sum of line segments. The desired differential form for such a geometry is given by
\begin{align}\label{Omega4alpha}
	\omega_4^{\alpha'} \!= \dd \log \frac{\sin(\pi\alpha' X_{13})}{\sin(\pi \alpha' (c_{13}-X_{13}))} = m_4^{\alpha'} \dd X_{13}.
\end{align}
Using Euler's infinite product formula for the sine function, this can be rewritten as
\begin{align}
	\omega_4^{\alpha'} \!= \sum_{k\in\mathbb{Z}} \dd \log \frac{X_{13}+k/\alpha'}{X_{13}-c_{13}+k/\alpha'}.
\end{align}
Comparing to \eqref{eq:Omega-line-segment}, we recognize this as a sum of canonical forms of an infinite number of line segments. The stringy geometry we are looking for is therefore
\begin{align}\label{defA4alpha}
    \mathcal{A}_4^{\alpha'}\coloneqq\bigcup_{k\in\mathbb{Z}} \{\frac{k}{\alpha'}\leq X_{13} \leq \frac{k}{\alpha'}+ c_{13}\},
\end{align}
whose canonical form is $\Omega(\mathcal{A}_4^{\alpha'})\!= \omega_4^{\alpha'} $. We note that $\mathcal{A}_4^{\alpha'}$ corresponds to the union of all translations of the ABHY associahedron $\mathcal{A}_4$ onto points of the one-dimensional lattice $1/\alpha'\;\mathbb{Z}$\footnote{We implicitly assume that $c_{13}<1/\alpha'$, as otherwise the line segments in $\smash{\mathcal{A}_4^{\alpha'}}$ will overlap. In this case we can still make sense of $\smash{\mathcal{A}_4^{\alpha'}}$ and its canonical form as a \emph{weighted positive geometry} \cite{Dian:2022tpf}. In the edge case where $c_{13}=1/\alpha'$ the geometry reduces to the entire real line, which has a vanishing canonical form. This gives a geometric interpretation to a `stringy hidden zero' of $m_4^{\alpha'}$, which can be derived from monodromy relations \cite{DAdda:1971wcy}.}.
The geometries of $\mathcal{A}_4^{\alpha'}$ and $\mathcal{A}_4$ are depicted in figure \hyperref[fig:A4-all]{1(b)} and \hyperref[fig:A4-all]{1(a)}, respectively.  

The above construction can be extended to any multiplicity $n$. We start from the ABHY associahedron $\mathcal{A}_n$ in an $(n-3)$-dimensional space of $X_{ij}$ variables, and translate it by a hypercubic lattice
\begin{align}\label{lattShift}
    \mathcal{A}_n^{\alpha'} \coloneqq \mathcal{A}_n + \frac{1}{\alpha'}\mathbb{Z}^{n-3}.
\end{align}
We refer to the resulting geometry $\mathcal{A}_n^{\alpha'}$ as an \emph{associahedral grid}. Its canonical form encodes the $n$-point diagonal matrix elements $m_n^{\alpha'}\!(\mathds{1}|\mathds{1})$ of the inverse KLT kernel via
\begin{align}\label{eq:canonical-form-An-string}
     \omega_n^{\alpha'} \equiv \Omega(\mathcal{A}_n^{\alpha'}) = m_n^{\alpha'} \dd^{n-3}X\,.
\end{align}
In the limit $\alpha' \to 0$ the associahedral grid $\mathcal{A}_n^{\alpha'}$ reduces to the field theory ABHY associahedron $\mathcal{A}_n$. This can be seen directly from the definition \eqref{lattShift} where all lattice sites involving $\alpha'$ move off to infinity.

To motivate the statement \eqref{eq:canonical-form-An-string}, we note that the associahedral grid can be triangulated via a \emph{geometric recursion}. Schematically, we decompose the geometry as
\begin{align}\label{eq:An-recursion}
    \mathcal{A}_n^{\alpha'} = \bigcup_{i=4}^n \mathcal{A}_4^{\alpha'}\times \mathcal{A}_{i-1}^{\alpha'}\times\mathcal{A}_{n-i+3}^{\alpha'},
\end{align}
where $\mathcal{A}_3^{\alpha'}$ is defined to be just a point. Iterating this recursion allows us to build up $\mathcal{A}_n^{\alpha'}$ in terms of sums of products of $\mathcal{A}_4^{\alpha'}$. We derive this relation in the \hyperref[sec:app]{End Matter} from an analogous relation for the ABHY associahedron $\mathcal{A}_n$. Using \eqref{eq:canonical-form-An-string}, the geometric recursion \eqref{eq:An-recursion} induces a recursion relation for the stringy matrix elements $m_n^{\alpha'}$, which we can write as
\begin{align}\label{eq:geometric-recursion-string}
    \hspace{-0.25cm} m_n^{\alpha'}\!=\!\sum_{i=4}^n \!\left. m^{\alpha'}_4\!\!(123\,i) [ m_{L_i}^{\alpha'}\!(2\dots i) m_{R_i}^{\alpha'}\!(i\dots 2) ] \!\right\vert_{X_{2j} \to X_{2j}-X_{2i}}\!,
\end{align}
where $L_i = i{-}1$, $R_i = n{-}i{+}3$ and the expressions in brackets are evaluated on shifted kinematics. 
Conversely, we find that the validity of \eqref{eq:canonical-form-An-string} can be derived from the recursion relation \eqref{eq:geometric-recursion-string}, which we have verified explicitly for $n{\leq} 10$ and conjecture to hold for all $n$.

\begin{figure}[t]
    \centering
    \includegraphics[width=\linewidth]{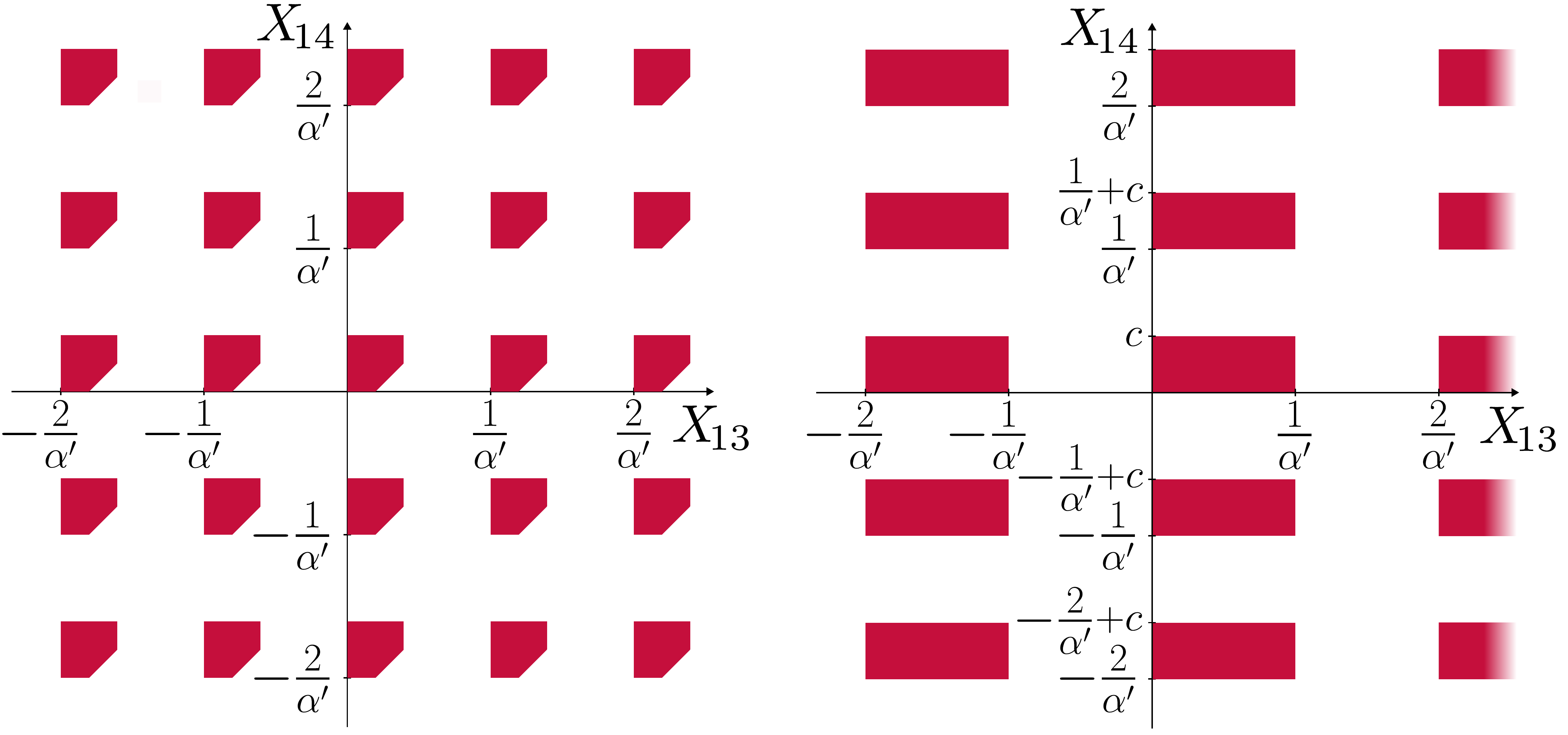}
    \caption{Associahedral grid for $m_5^{\alpha'}\!(\mathds{1}|\mathds{1})$ \eqref{invKLTex} (left), and the off-diagonal element $m_5^{\alpha'}\!(\mathds{1}|13452)$ \eqref{exKLToffdiag} (right), where we use $c=c_{14}+c_{24}$.}
    \label{fig:A5-grid}
\end{figure}

Let us consider the first non-trivial example of the geometric recursion \eqref{eq:An-recursion} for $n=5$,
\begin{align}
     \hspace{-0.25cm} \mathcal{A}_5^{\alpha'}= &\{ 0\leq X_{14} \leq c_{14} \}^{\alpha'} \!\!\times\! \{ 0\leq X_{13} \leq c_{13}+X_{14} \}^{\alpha'} \label{eq:A5-string-triangulation} \\
    \cup  &\{c_{14}\leq X_{14} \leq c_{14}+c_{24}\}^{\alpha'}\!\!\times\! \{ 0\leq X_{13} \leq c_{13}+c_{14} \}^{\alpha'}\!, \nonumber
\end{align}
where we use the notation $\{a\leq x\leq b\}^{\alpha'} = \{a\leq x\leq b\}$ $+1/\alpha'\;\mathbb{Z}$ for the `stringy line segment', which we interpret as instances of $\mathcal{A}_4^{\alpha'}$. Expanding the product, we see that \eqref{eq:A5-string-triangulation} triangulates the associahedral grid $\mathcal{A}_5^{\alpha'}$ depicted in figure \ref{fig:A5-grid} (left). We ignore terms with a vanishing canonical form in this expansion, which is explained in more detail in the \hyperref[sec:app]{End Matter}. Removing the $\alpha'$ labels from the above formula gives a triangulation of $\mathcal{A}_5$ (\textit{cf.} equation \eqref{eq:A5-chamber-fiber}).

Since $\mathcal{A}_n$ has boundaries exactly when some $X_{ij}$ goes to zero, it is easy to see from \eqref{lattShift} that $\smash{\mathcal{A}_n^{\alpha'}}$ has boundaries when some $X_{ij} = k/\alpha'$ for some $k\in\mathbb{Z}$. This is in agreement with the expected pole structure \eqref{periodPoles}. The associahedral grid is arguably the \emph{simplest} geometry which satisfies this condition. Still, it is far from trivial that the canonical forms of these associahedra sum to give exactly the right function $m_n^{\alpha'}$, including all the contact terms.

We have argued that the resonance structure \eqref{periodPoles} of $m_n^{\alpha'}$ is reflected in the periodicity of the associahedral grid. Alternatively, it can be captured by a single ABHY associahedron with kinematic space compactified onto a torus, hinting at a possible connection to \cite{Frost:2018djd}, where the inverse KLT kernel is interpreted via toric varieties.

The geometry of $\mathcal{A}_n^{\alpha'}$ further suggests an interesting rewriting of the stringy matrix elements $m_n^{\alpha'}$ as an infinite sum of shifted field theory amplitudes. For example, at four points we have the identity
\begin{align}\label{m4shft}
    m_4^{\alpha'} \!= \sum_{k\in\mathbb{Z}} \left(\! \frac{1}{X_{13}-k/\alpha'} + \frac{1}{X_{24}+k/\alpha'}\!\right)\!.
\end{align}
Curiously, only the sum of the entire shifted amplitude is well defined, as divergent parts between individual terms cancel. This phenomenon likewise occurs at higher multiplicities where it is crucial for the emergence of the infinite tower of contact terms in $m_n^{\alpha'}$.

\subparagraph{III. Off-Diagonal Matrix Elements.}
The associahedral grid $\mathcal{A}_n^{\alpha'}$, which captures the diagonal elements of the inverse KLT kernel $m_n^{\alpha'}\!(\mathds{1}|\mathds{1})$, can also be used to find geometries for off-diagonal elements $m_n^{\alpha'}\!(\sigma|\rho)$. 
We recall that off-diagonal elements can be written as products of diagonal elements times factors of $1/\sin(\pi\alpha'X_{ij})$. Thus, if we find a geometry for some factor $1/\sin(\pi\alpha'X_{ij})$, we can multiply it with products of $\mathcal{A}_n^{\alpha'}$ to find a geometry for any off-diagonal element\footnote{Depending on the basis of $X_{ij}$s and $c_{ij}$s we choose for our associahedron, the elements in this product might overlap, in which case we again need to make use of the formalism of weighted positive geometries.}. 
We find that the appropriate differential form for the inverse sine factor is
\begin{align}\label{dformSine}
\begin{split}
	\dd\log\tan(\pi\alpha' X_{ij}/2) & = \frac{\dd X_{ij}}{\sin(\pi\alpha' X_{ij})}
	\\ &= \sum_{k\in\mathbb{Z}} \dd\log\frac{X_{ij}-2k/\alpha'}{X_{ij}-(2k+1)/\alpha'}\,.
\end{split}
\end{align}
Again, this can be interpreted as the canonical form of an infinite union of line segments $\bigcup_{k\in\mathbb{Z}} \{2k/\alpha \leq X_{ij}\leq (2k+1)/\alpha'\}$, which we depict in figure \hyperref[fig:A4-all]{1(c)}. 

Combining this result with the geometry $\mathcal{A}_n^{\alpha'}$ for the stringy $\trp$ amplitudes, we find a geometry $\mathcal{A}_{\alpha'}(\alpha|\beta)$ for \emph{any} off-diagonal element. A simple 5-point example is depicted in figure \ref{fig:A5-grid} (right), where $m_5^{\alpha'}\!(\mathds{1}|13452)$ \eqref{exKLToffdiag} is constructed from the product of a single inverse sine factor \eqref{dformSine} and the 4-point geometry \eqref{Omega4alpha}.

\subparagraph{IV. Pions and Mixed Amplitudes.} 
It was recently shown in \cite{Bartsch:2025loa} that the inverse KLT kernel $m_n^{\alpha'}$ not only contains $\trp$ amplitudes in its $\alpha'\to 0$ limit, but also NLSM and mixed $\phi$/$\pi$ amplitudes via kinematic $\alpha'$-shifts. In particular, rescaling $\alpha'\to\alpha'/2$ and shifting certain $X_{ij}\mapsto X_{ij}\pm1/\alpha'$, any of these amplitudes can be obtained from $m_n^{\alpha'}$ in the $\alpha'\to 0$ limit. 

These $\alpha'$-shifts can be interpreted geometrically. For the associahedral grid $\mathcal{A}_n^{\alpha'}$, the rescaling $\alpha'\to\alpha'/2$ stretches the grid and effectively removes all lattice points at odd integer multiples of $1/\alpha'$. Assuming the $\alpha'$-shift preserves the $c_{ij}$ variables in our kinematic basis, it corresponds to a simple translation of the entire lattice. The result is a subgrid of ABHY associahedra, whose canonical form yields stringy pion or mixed amplitudes.
\begin{figure}[t]
    \centering
    \includegraphics[width=\linewidth]{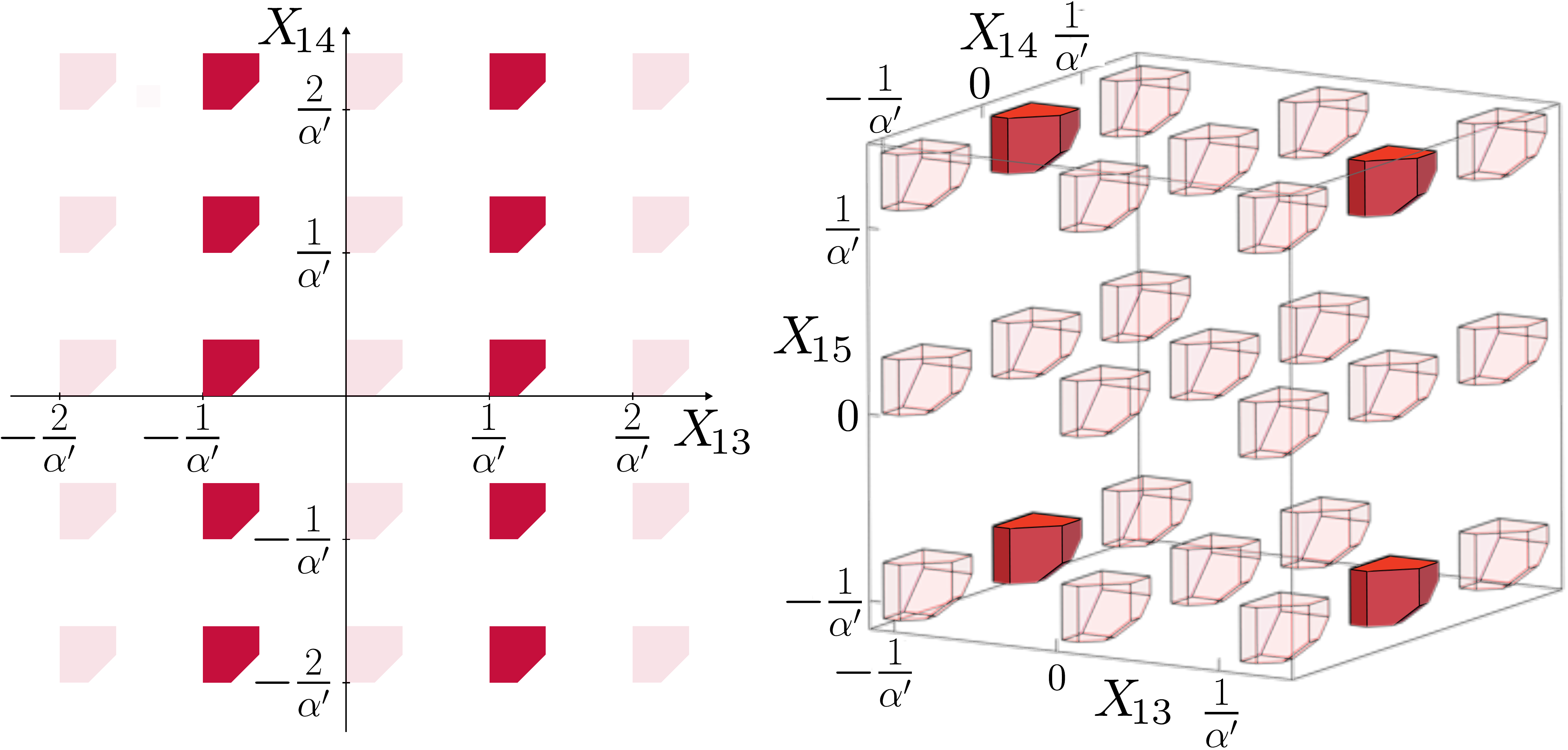}
    \caption{Associahedral subgrid for the stringy mixed amplitude $M_5^{1\pi,\alpha'}\!(\pi\phi\phi\phi\phi)$ \eqref{M1pialpha} (left), and stringy NLSM at 6-points \eqref{NLSM6alpha} (right). }
    \label{fig:n56-sub-grid}
\end{figure}

As an example, let us consider $\mathcal{A}_4^{\alpha'}$ as shown in figure \hyperref[fig:A4-all]{1(b)}. Rescaling $\alpha'\to \alpha'/2$ leaves us with a lattice of ABHY associahedra positioned at $2k/\alpha'$, $k\in\mathbb{Z}$. We now shift $X_{13}\to X_{13}+1/\alpha'$, which translates the lattice over by one unit of $1/\alpha'$ (this automatically shifts $X_{24}\to X_{24}-1/\alpha'$, since in our basis $X_{24}=c_{13}-X_{13}$) and end up with the grid depicted in figure \hyperref[fig:A4-all]{1(d)}. This infinite subgrid, which we denote $\mathcal{A}_4^{\text{NLSM}, \alpha'}$, has a canonical form of 
\begin{align}\label{NLSM4dform}
\begin{split}
	&\Omega(\mathcal{A}_4^{\text{NLSM}, \alpha'})=\dd\log\frac{\cos(\pi\alpha'X_{13})}{\cos(\pi\alpha'(X_{13}-c_{13}))} \\
	&= -\left(\tan(\pi\alpha'X_{13}) + \tan (\pi\alpha' X_{24})\right)\dd X_{13}\,.
\end{split}
\end{align}

Let us investigate the $\alpha'\to0$ limit of this function. We first look at the line segment $\{1/\alpha'\leq X_{13} \leq 1/\alpha' + c_{13}\}$ of the infinite grid in $\mathcal{A}_4^{\text{NLSM},\alpha'}$ closest to the origin. Its canonical form is equivalent to the $\delta$-shifted $\trp$ amplitude of \cite{Arkani-Hamed:2024nhp}, where all $X_{\text{oo}}\to X_{\text{oo}}+\delta,\,X_{\text{ee}} \to X_{\text{ee}}-\delta$, with $o/e$ referring to odd/even indices. Identifying $\delta = 1/\alpha'$, the $\delta\to \infty$ (or, equivalently, $\alpha'\to 0$) behavior of this shifted amplitude is
\begin{align}
	\frac{1}{X_{13}+1/\alpha'} + \frac{1}{X_{24}-1/\alpha'} \propto \alpha'^2 (-X_{13}-X_{24}) + \mathcal{O}(\alpha'^3),
\end{align}
which correctly reproduces the 4-point NLSM amplitude $A_4^{\text{NLSM}}=-X_{13}-X_{24}$. All other line segments $\{k/\alpha'\leq X_{13}\leq k/\alpha'+c_{13}\}$ in the infinite grid contribute the same NLSM amplitude with a scaling of $1/k^2$, which sum to give a prefactor $\pi^2/4$. Thus, up to a constant, the $\alpha'\to0$ limit of $\Omega(\mathcal{A}_4^{\text{NLSM},\alpha'})$ is equivalent to the $\delta \to \infty$ limit of the $\delta$-shifted $\trp$ amplitude. An analogous story holds for the 6-point NLSM amplitude, the infinite subgrid of which is depicted in figure \ref{fig:n56-sub-grid} (right). The canonical form of this geometry is
\begin{align}
    \Omega(\mathcal{A}_6^{\text{NLSM},\alpha'}) = A_6^{\alpha'} \dd X_{13}\dd X_{14}\dd X_{15},
\end{align}
and encodes the stringy pion function \cite{Bartsch:2025loa}
\begin{align}\label{NLSM6alpha}
    A_6^{\alpha'} = \frac{1}{2}\frac{(\tau_{13}+\tau_{24})(\tau_{46}+\tau_{15})}{\tau_{14}}-\tau_{13}-\frac{1}{3}\tau_{13}\tau_{35}\tau_{15} + \text{cyc.},
\end{align}
where we use the shorthand $\tau_{ij} = \tan(\frac{\pi}{2}\alpha' X_{ij})$.

Furthermore, associahedral subgrids can describe mixed $\phi/\pi$-functions which are not immediately accessible through an $\alpha'$-shift of the inverse KLT kernel or a $\delta$-shift of the $\trp$ amplitude. For example, at five points we can define a stringy `one-pion' function,
\begin{align}\label{M1pialpha}
\begin{split}
    M_5^{1\pi,\alpha'} &= -\frac{\tau_{14}+\tau_{25}}{\tau_{24}} - \frac{\tau_{13}+\tau_{25}}{\tau_{35}} - \frac{\tau_{13}+\tau_{24}}{\tau_{14}} + 2 \\
    &+ \tau_{13}(\tau_{14}+\tau_{35}) + \tau_{25}(\tau_{24}+\tau_{35}),
\end{split}
\end{align}
whose associahedral grid is shown in figure \ref{fig:n56-sub-grid} (left). It can be shown that this grid cannot be obtained from the associahedral grid of $m_5^{\alpha'}$ in figure \ref{fig:A5-grid} (left) just by rescaling and shifting along the axes using a single $\alpha'$-shift. However, the geometry of the subgrid suggests that \eqref{M1pialpha} \emph{can} be obtained as linear combination of two distinctly $\alpha'$-shifted matrix elements $m_5^{\alpha'}$. This illustrates that the associahedral grid allows to discover new classes of functions connected to the inverse KLT kernel beyond those previously known in the literature, notably functions involving an odd number of pions. For $\alpha'\to 0$ \eqref{M1pialpha} reduces to the amplitude $M_5^{1\pi}(\pi\phi\phi\phi\phi)$ appearing as the leading-order contribution to the semi-abelianized disk integral $Z_{1234}(\mathds{1})$ of \cite{Carrasco:2016ygv}.

We emphasize that, although we have found a positive geometry for \emph{stringy} NLSM and $\phi/\pi$ amplitudes, the search for a positive geometry for their field theory counterparts continues. Considering the $\alpha'\to0$ limit of our subgrid interpretation of stringy NLSM amplitudes suggests to look for a suitable interpretation of `geometry at infinity' exhibited by NLSM amplitudes. 

\section{Conclusion and Outlook}
We have introduced a novel positive geometry, the associahedral grid $\mathcal{A}_n^{\alpha'}$, which consists of an infinite lattice of ABHY associahedra. It unifies stringy amplitudes for $\trp$ theory, the inverse KLT kernel, NLSM, and mixed $\phi$/$\pi$ amplitudes, capturing the full $\alpha'$-dependence in a simple geometric framework. This result demonstrates how positive geometries can be extended beyond rational functions to accommodate genuinely stringy structures. Our results further suggest a natural interpretation of the $\delta$-shift as corresponding to some subgrid of $\mathcal{A}_n^{\alpha'}$.

We have illustrated that it is possible to find a positive geometry for stringy NLSM amplitudes, whereas a geometric description of field theory NLSM remains unknown. This highlights both the limitations and the promise of the positive geometry framework, and naturally raises the prospects of extending this construction to other theories. The connection to stringy special Galileon amplitudes, related to the stringy NLSM amplitudes presented here via abelianization \cite{Bartsch:2025loa}, is of the most immediate interest. A more thorough understanding of these amplitudes may help elucidate how geometric and combinatoric methods can be used for permutation-invariant amplitudes, and could be a concrete step towards a geometric description of gravity amplitudes.

\medskip
{\it Acknowledgements}: We thank Jaroslav Trnka for useful discussions and comments on the draft. This work is supported by GA\v{C}R 24-11722S and OP JAK CZ.02.01.01/00/22\_008/0004632.
\vspace{-0.55cm}
\bibliography{mainbib}
\bibliographystyle{apsrev4-1}

\onecolumngrid

\section{End Matter}\label{sec:app}
\noindent We will derive \emph{geometric recursion relations} for $\trp$ amplitudes from the ABHY associahedron, which we conjecture also to hold for diagonal matrix elements of the inverse string theory KLT kernel. The validity of these recursion relations for $m_n^{\alpha'}$ serves as the main argument for the validity of the associahedral grid picture we develop in this letter. The derivation of these recursion relations is inspired by recent progress in understanding loop integrands in planar $\mathcal{N}=4$ SYM \cite{Ferro:2023qdp, Ferro:2024vwn} and ABJM theory \cite{He:2023rou, Lukowski:2023nnf} via a geometric procedure where the loop degrees of freedom are interpreted as \emph{fibrations} over tree-level \emph{chambers}. 

Starting from the $(n-3)$-dimensional object $\mathcal{A}_n$ in the standard $(X_{13},X_{14},\ldots,X_{1n-1})$ basis, we can project out one degree of freedom, say $X_{13}$. The image of this projection will be a lower-dimensional associahedron $\mathcal{A}_{n-1}$. We can reconstruct the full geometry by attaching a \emph{fiber} (a line segment) to each point in the image of the projection. The mathematical description of this line segment will differ for different points in the image. This suggests a triangulation of $\mathcal{A}_{n-1}$ into several \emph{chambers} $\mathfrak{c}_i$, which are defined such that the mathematical description of the fiber $f_i$ is equivalent for all points inside a chamber. This allows us to write the full $n$-point ABHY associahedron as
\begin{align}
	\mathcal{A}_n=\bigcup_i \mathfrak{c}_i\times f_i\,,
\end{align}
where the union is over all chambers. This leads to a formula for $m_n$ as
\begin{align}
	m_n \dd X_{13}\wedge\cdots\wedge \dd X_{1n-1} =\sum_i \Omega(\mathfrak{c}_i)\wedge \Omega(f_i)\,.
\end{align}
In our case the chambers are the projection of boundaries of $\mathcal{A}_n$ given by $X_{2i}=0$, and hence are products of two lower-point associahedra. The fibers take the form
\begin{align}
    f_i = \{0\leq X_{13} \leq C_i+X_{1i}\},\qquad C_i=\sum_{j=3}^{i-1} c_{1j}.
\end{align} 
This leads to a triangulation of $\mathcal{A}_n$ which looks identical to equation \eqref{eq:An-recursion} if we drop the $\alpha'$. 

\begin{figure}
    \centering
    \includegraphics[width=0.65\linewidth]{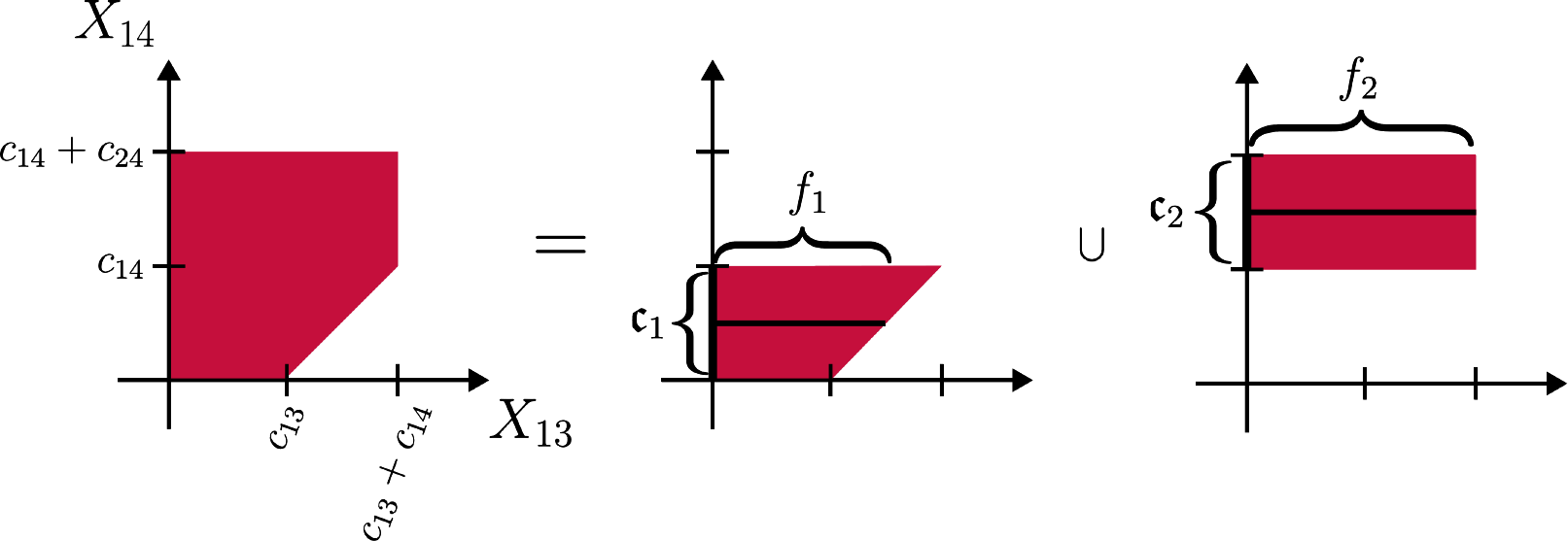}
    \caption{The 5-point associahedron $\mathcal{A}_5$ is triangulated by the chambers and fibers in equation \eqref{eq:A5-chamber-fiber}.}
    \label{fig:chamber-fiber}
\end{figure}

The first non-trivial example occurs when $n=5$. For our choice of basis and projection, this leads to a triangulation of $\mathcal{A}_5$ as 
\begin{align}\label{eq:A5-chamber-fiber}
	\mathcal{A}_5= \mathfrak{c}_1\times f_1 \cup \mathfrak{c}_2\times f_2 = &\{0 \leq X_{14} \leq c_{14}\}\times \{0 \leq X_{13}\leq c_{13}+X_{14}\} \\\notag \cup &\{c_{14} \leq X_{14} \leq c_{14}+c_{24}\} \times \{0 \leq X_{13} \leq c_{13}+c_{14}\}\,,
\end{align}
which is illustrated in figure \ref{fig:chamber-fiber}.
When written in terms of $X_{ij}$ variables, this gives an expression for the amplitude as
\begin{align}
	m_5 &= \left(\frac{1}{X_{14}}+\frac{1}{X_{25}-X_{24}}\right)\!\!\left(\frac{1}{X_{13}}+\frac{1}{X_{24}}\right) + \left(\frac{1}{X_{35}}+\frac{1}{X_{24}-X_{25}}\right)\!\!\left(\frac{1}{X_{13}}+\frac{1}{X_{25}}\right).
\end{align}
In general, it can be shown that this procedure is equivalent to the following recursion relation for $m_n$:
\begin{align}\label{eq:trp-recursion}
	m_n= \sum_{i=4}^n \underbrace{m(1,2,3,i)}_{f_i} \underbrace{\left.m(2,3,\ldots,i)m(i,i+1,\ldots,n,1,2)\right|_{X_{2j}\to X_{2j}-X_{2i}}}_{\mathfrak{c}_i}.
\end{align}
One should take care not to confuse the arguments of the $\trp$ amplitudes in this recursion with the double color ordering of BAS amplitudes. Here we instead interpret the scattering amplitudes as functions of the labels of the external particles, \emph{e.g.} $m(1,2,3,i)=1/X_{13}+1/X_{2i}$. This formula provides an efficient recursion relation for $m_n$ with the interpretation that we can get higher-point amplitudes by \emph{fibrating} particles over lower-point amplitudes. The chambers have spurious poles of the form $1/(X_{2i}-X_{2j})$, which cancel out in the total sum\footnote{It is worth pointing out that these spurious poles also appear in \cite{Jones:2025rbv}, which could be an indication that the formulae presented there can be given a similar geometric interpretation.}. Iterating the recursion allows us to build up $\mathcal{A}_n$ entirely as a sum of products of line segments $\mathcal{A}_4$.

To extend this discussion to the associahedral grid we first define the stringy chambers $\mathfrak{c}_i^{\alpha'}=\mathfrak{c}_i+\mathbb{Z}^{n-4}/\alpha'$, which triangulate the image of $\mathcal{A}_n^{\alpha'}$ after projecting through $X_{13}$. We consider a fixed element $\mathfrak{c}_i^*\equiv \mathfrak{c}_i+ (a_4,\ldots,a_{n-1})/\alpha'$ of $\mathfrak{c}_i^{\alpha'}$, $a_j\in\mathbb{Z}$. The part of the associahedral grid which projects onto this chamber is given by
\begin{align}\label{eq:1}
    \mathfrak{c}_i^*\times\bigcup_{k\in\mathbb{Z}}\{ \frac{k}{\alpha'}\leq X_{13}\leq C_i + X_{1i} + \frac{k-a_i}{\alpha'} \}.
\end{align}
The canonical form of the fiber attached to $\mathfrak{c}_i^*$ is equivalent to that of the stringy line segment
\begin{align}
    f_i^{\alpha'}=\bigcup_{k\in\mathbb{Z}} \{\frac{k}{\alpha'}\leq X_{13} \leq \frac{k}{\alpha'}+C_i+X_{1i}\}.
\end{align}
This equivalence is obtained by writing
\begin{align}
    \Omega(f_i^{\alpha'})=\sum_{k\in\mathbb{Z}}\Omega\big(\{ \frac{k}{\alpha'}\leq X_{13}\leq C_i + X_{1i} + \frac{k-a_i}{\alpha'} \}\big)+ \sum_{k\in\mathbb{Z}}\Omega\big(\{ \frac{k}{\alpha'} \leq X_{13} \leq \frac{k-a_i}{\alpha'} \}\big).
\end{align}
It is easy to see that the second part of this expression vanishes for any integer $a_i$. The importance of this statement comes from the observation that $f_i^{\alpha'}$ does not depend on the position of $\mathfrak{c}_i^*$ in the lattice $\mathbb{Z}^{n-4}/\alpha'$, and hence no infinite summation needs to be solved to find the canonical form of $\mathcal{A}_n^{\alpha'}$. Instead, we arrive at the result that
\begin{align}
    \Omega(\mathcal{A}_n^{\alpha'})=\sum_i\Omega(\mathfrak{c}_i^{\alpha'})\wedge\Omega(f_i^{\alpha'}).
\end{align}
This implies that the recursion relation \eqref{eq:trp-recursion} induces a recursion relation for $m_n^{\alpha'}$:
\begin{align}\label{eq:trp-recursion-stringy}
	m_n^{\alpha'}= \sum_{i=4}^n m^{\alpha'}(1,2,3,i) \left.\big[m^{\alpha'}(2,3,\ldots,i)m^{\alpha'}(i,i+1,\ldots,n,1,2)\big]\right|_{X_{2j}\to X_{2j}-X_{2i}}.
\end{align}
The spurious poles now generate contact terms in $m_n^{\alpha'}$ through the trigonometric relation
\begin{align}
	\frac{1}{\tan(\pi\alpha'(X_{2j}-X_{2i}))} = \frac{1+\tan(\pi\alpha'X_{2i})\tan(\pi\alpha'X_{2j})}{\tan(\pi\alpha'X_{2i})-\tan(\pi\alpha'X_{2j})}\,.
\end{align} 
It is a small miracle that all the spurious poles in \eqref{eq:trp-recursion-stringy} conspire to give \emph{exactly} all the contact terms of $m_n^{\alpha'}$. We have verified this statement explicitly up to $n=10$.

\end{document}